\documentstyle[prl,aps,multicol]{revtex}

\def\beq{\begin{equation}}
\def\eeq{\end{equation}}

\def\6{\langle }
\def\9{\rangle }

\begin{document}

\title{Comment on `Counterfactual entanglement and non-local correlations in
separable states'.}

\author{Daniel R. Terno \thanks {e-mail: terno@physics.technion.ac.il} 
\\{\sl Department of Physics,
Technion--- Israel Institute of Technology, 32000 Haifa, Israel}}

\maketitle
\begin{abstract}

The arguments of Cohen [Phys. Rev. A {\bf 60}, 80 (1999)] against the
`ignorance interpretation'
of mixed states are questioned. The physical arguments are shown to be inconsistent and
the supporting example illustrates the opposite of the original statement. The
operational difference between two possible definitions of mixed states is
exposed and the inadequacy of one of them is stressed.
$ $

\end{abstract}
\begin{multicols}{2}

In a recent paper~\cite{co} Cohen constructs an example of 
counterfactual entanglement and also deals with
 two possible scenarios that lead to the appearence of mixed states. 
 These states arise either from tracing out unavailable degrees of freedom (`ancilla 
interpretation') or from the actual mixing of pure states (`ignorance interpretation').
Cohen states that the ignorance interpretation is `unsatisfactory'.  The abovementioned
counterfactual entanglement and the arguments against the ignorance interpretation allow
to question whether it is `appropriate to label weighted sums of projections on product states
as ``separable mixed states'''. 

The relevance of  counterfactual entanglement to the
 physical meaning of separable states and its deeper implications on the non-locality 
problems will not be discussed here. 
 However, while some of Cohen's arguments against the ignorence interpretation are of 
philosophical nature and their 
acceptance or refutal is a matter of  personal opinion, several claims can be rigorously 
analyzed and  are demonstrably wrong. It should be stressed, however, that while their 
refutal may weaken the case against the existence of separable states, it by no 
means undermines the validity of Cohen's example of the countrfactual entanglement. 
Moreover, it should be noted that some
of Cohen's arguments against the ignorance interpretation of  mixed states follow a similar 
discussion in the book of d'Espagnat \cite{de}.

One of the  arguments of~\cite{co} is that `in some cases different
statistical mixtures of pure states may appear to be representable by the
same density matrix but may nevertheless be experimentally distinguishable'.
An example which is meant to illustrate this argument  is a comparison
between two large sets of spin-$\frac{1}{2}$ particles. In the first set exactly $N$ out of $2N$  
particles are prepared in spin-up state in $z$ direction and the other half is
prepared spin-down in $z$ direction. Another set is prepared according to
the same recipe, but along the $x$-axis.  As shown below the example illustrates exactly the opposite
of Cohen's claim.

We begin by noting that the density operator that we ascribe to a state 
depends not only on the preparation procedure, but also on the
experimental techniques that can be used and on the number of systems available to
the test. Moreover, either there is a finite probability to distinguish
between the states, or they have the same density matrix (again, the
confidence level and the possibility of the procedure itself depend on the
quality of experimental techniques). 

Let us consider the simplest
distinguishability criterion: probability of error. It deals with
two states whose density matrices $\rho_1$ and $\rho_2$ are known. An observer is given one 
of them and is allowed to perform any lawful quantum operation. At the end,
the observer should give an unambiguous answer which state it was. The
probability to err is  \cite{hel}
\beq
P_E=\frac{1}{2}-\frac{1}{4} {\rm Tr} |\rho_1-\rho_2|,
\eeq
and it depends only on the density matrices. In particular, identity of density matrices is equivalent 
to total indistinguishability of the states and distinguishable states cannot `appear to be 
representable by the same density matrix'. 

Now let us examine Cohen's example.  We suppose that the detectors are ideal 
and consider two
possible ways to identify the states. First, we analyze the  experiment where particles are tested 
individually along the same axis (say, $z$-axis). If we have the first set, we 
should get exactly $N$ up and $N$ down results, 
while the probabilities of these  outcomes for the second set are distributed binomially. It is easy to
see that the probaility of error in the identification is
\beq
P_E=\frac{1}{2}\times\frac{1}{2^{2N}} \frac{(2N)!}{(N!)^2}\simeq\frac{1}{2\sqrt{\pi N}}, \label{di}
\eeq

On the other hand, multiparticle measurements not only improve the distinguishability,
but also highlight the differences between preparation procedures. If the particles
can be considered as distinguishable (like qubits in the register of a quantum computer), then no 
symmetrisation is needed. The  preparations can be represented by two pure staes $\psi_1$ and $\psi_2$,
which are composed from direct products of  
eigenstates of $\sigma_z$ and  eigenstates of $\sigma_x$ respectively. 
If we take into account that the overlap between any  eigenstate of
$\sigma_z$ and any eigenstate of  $\sigma_x$ is $|\6 u_i|v_j\9|=\frac{1}{\sqrt{2}}$,
the overlap between $\psi_1$ and $\psi_2$ is
\beq
|\6\psi_1|\psi_2\9|=\frac{1}{2^N}. 
\eeq

If the particles are linearly polarized photons and polarization is the only
 degree of freedom, then the correct description of  both preparations is given by
 symmetric states in the Fock space. If $\hat{a}^{\dag}$ and
$\hat{b}^{\dag}$ are creation operators in two perpendicular directions, and
$|0,0\9$ is the vacuum state, then the first state is given by
\beq
\psi_1=(\hat{a}^{\dag}\hat{b}^{\dag})^N |0,0\9/N!=|N,N\9,
\eeq
while the second state is
\beq
\psi_2=(\hat{a}^{\dag}+\hat{b}^{\dag} )^N (-\hat{a}^{\dag}+\hat{b}^{\dag})^N|0,0\9/(2^N N!).
\eeq
Moreover, it is easy to see that these states are in fact almost (or even exactly) {\em orthogonal}. 
 Using formulas 0.157 of \cite{gr}, we find that their
overlap $|\6\psi_1|\psi_2\9|$ for even $N$ is
\beq
|\6\psi_1|\psi_2\9|=\frac{1}{2^N}\frac{1}{N!}\left|\sum_{k=0}^N (-1)^k
\left(\begin{array}{c} N \\k \end{array} \right)^2\right|=\frac{1}{2^N ((N/2)!)^2} ,
\eeq
while for odd $N$ the states are orthogonal,
\beq
|\6\psi_1|\psi_2\9|=0.
\eeq
(This happens, e.g., for $N=1$, giving a spin-1 triplet state). To compare  two modes of 
 investigation we  note that for two pure states $\psi_1$ and $\psi_2$
\beq
P_E(\psi_1,\psi_2)= \frac{1}{2}(1-\sqrt{1-|\6\psi_1|\psi_2\9|^2}).
\eeq
Thus for the distinguishable particles we have 
\beq
P_E=\frac{1}{2^{2N+2}}.
\eeq
For the photons we have
\beq
P_E\simeq|\6\psi_1|\psi_2\9|^2/4=
\frac{1}{(2\pi e)^2}\left(\frac{e}{N}\right)^{2N+2}
\eeq
for small overlaps (large $N$)  when $N$ is even, and $P_E=0$ (exact distinguishability) for odd $N$.
Thus we see that while we ought to agree with Cohen that the states are distinguishable, it is 
impossible to ascribe to them the same density matrix (especially if they are orthogonal).

To refine the above analysis we  clarify what is exactly meant by an  `ignorance
mixture'. Two definitions are found in literature.  Namely, the density operator is defined
either as
\beq
\rho^{(1)}=\sum_i N_i |u_i\9\6u_i|/N,~~ N=\sum_i N_i,
\eeq
where $N_i$'s are exact numbers (`type-1 preparation' ,\cite{de}), or
\beq
\rho^{(2)}=\sum_i p_i |u_i\9\6u_i|,~~ \sum_i p_i=1,
\eeq
where $p_i$'s are probabilities  to have states $|u_i\9$ (`type-2 preparation',\cite{pe}). It is not stated 
explicitely in~\cite{co} to which definition the author subscribes (however all examples are 
of the type-1). 

Type-1 and type-2 preparations of `the same state' $\rho$ are not equivalent. 
They may be
distinguishable (with finite probability) if more than one copy is available. 
Consider a preparation of type-1 of exactly $N$ spin-up and and $N$ spin-down particles
along  some known axis,  and a preparation of type-2 with $p_1=p_2=N/2N$ along the same axis. When 
 detectors are ideal and particles
are tested individually, the probability to err in their identification is given by Eq.~(\ref{di}). The
immediate consequence of this result is that the preparation-1 type state cannot, in general,
 reproduce the local statistics of the EPR experiment. This statistics is reproduced by the maximally
mixed state $\rho^{(2)}$. This result, together with Eq.~(\ref{di}) and above examples, implies
 that despite its appearence, the type-1 preparation is not described by the  maximally
mixed density matrix (the correct description depends on the exact details of the preparation and
may even  be given by a pure state).

The subtle dependence of the ascribed $\rho$ on the details of the preparation
and the observation procedures is further illustrated by the examples below. 
Consider again Cohen's example, but let us replace  type-1  preparations 
by those of type-2, with $p_i=N_i/N$. 

If only individual particles can be tested, then for both preparations the 
probabilities of the
oucomes are described by the same binomial distribution. Thus they are
 indistinguishable and the correct description of the states should be given
by mixed density matrices $\rho_1=\rho_2={\bf 1}/2$, where ${\bf 1}$ is 
a unit matrix. This result is independent from the
 number of available particles. 

In the case of distinguishable particles  the multiparticle state reduces to the direct product 
of the individual density matrices. Obviously, in this case it is also impossible to distinguish between
the preparations.

On the other hand, mixtures of different $2N$-boson states lead to a different conclusion.
 Since the expressions becomes cumbersome for large $N$, let us consider the
simplest case of two-particle states. Now we have
\beq
\rho_1=\frac{1}{4}|0,2\9\60,2|+\frac{1}{2}|1,1\9\61,1|+\frac{1}{4}|2,0\9\62,0|,
\eeq
and
\begin{eqnarray}
\rho_2 & = & \frac{1}{4}|\phi(0,2)\9\6\phi(0,2)|+\frac{1}{2}|\phi(1,1)\9\6\phi(1,1)|\\ \nonumber
& &+ \frac{1}{4}|\phi(2,0)\9\6\phi(2,0)|,
\end{eqnarray}
with
\beq
\phi(k,2-k)=\frac{1}{2 \sqrt{k!}\sqrt{(2-k)!}}
(\hat{a}^{\dag}+\hat{b}^{\dag} )^k (-\hat{a}^{\dag}+\hat{b}^{\dag})^{2-k}|0,0\9.
\eeq
A straightforward calculation gives
\beq
P_E=\frac{1}{2}-\frac{1}{4}\times \frac{1}{2}=\frac{3}{8},
\eeq
and there is a nonvanishing probability to distinguish between these two states.

It is also stated in \cite{co} that `given an EPR spin-singlet pair each separate particle can
be described by a mixed state, with the other particle then taking the role of ancilla. But if we
then assume that this mixed state can also be given an ignorance interpretation, inconsistencies
immediately arise'. We claim that no inconsistencies arise, as shown below.  The  local statistics 
of both EPR particles can be described by the maximally
mixed density matrix. A possible  interpretation of this local state is that it originates
from a type-2 preparation procedure, since type-1 is inconsistent with experiment. However, when 
the correlations between the particles are analyzed, the description of the complete 
system in terms of 
mixed states should be dropped, regardless of our interpretation of Bell's theorem or  
 philosophical attitudes, but simply because it is inconsistent with experiment.

To summarize, we see that the type-1 interpretation is unsuitable for the description of 
mixed states, but because of  reasons different from those that are given in \cite{co}. On the 
other hand, the type-2 realisation of mixed states leads to a consistent description of
physical systems and its predictions are identical to those obtained with the help of ancilla
 interpretation.

\bigskip\noindent{\bf Acknowledgments}\medskip

Discussions with Asher Peres and his help in the preparation of the manuscript
are gratefully acknowledged. Several of the above examples are extensions of the exercises 
from chapters 2 and 5 of his book \cite{pe}. I also thank Oliver Cohen for clarifying
important points of his article. This work is supported
 by a grant from the Technion Graduate School.

\end{multicols}

\begin{thebibliography}{99}
\bibitem{co} O. Cohen, Phys. Rev. A {\bf 60}, 80 (1999).
\bibitem{de} B. d'Espagnat, {\em Veiled reality} (Addison-Wesley, Reading MA, 1994), p.~83.
\bibitem{hel} C. W. Helstorm, {\em Quantum detection and estimation theory},
 (Academic Press, New York, 1976).
\bibitem{gr} I.S. Gradshteyn, I.M. Ryzhik, ed. by A. Jeffrey, {\em Table of integrals,
series, and products}, (Academic Press, San Diego, 1994).
\bibitem{pe} A. Peres, {\em Quantum mechanics: concepts and methods}, (Kluwer,
Dordrecht, 1993).
\end{thebibliography}
\end{document}